\documentclass{PoS}
\usepackage{gensymb}
\usepackage[makeroom]{cancel}
\def\be{\begin{equation}}
\def\ee{\end{equation}}
\newcommand \helicityAmp {\mathcal{M}^{m_{J_A}m_{J_B}m_{J_C}}}

\title{Heavy Quark Spin Symmetry Violating Hadronic Transitions of Higher Charmonia}

\ShortTitle{HQSS Violating Hadronic Transitions of Higher Charmonia}

\author{\speaker{Muhammad Naeem Anwar} \thanks{This work is supported by the National Natural Science Foundation of
                                               China (NSFC) through funds provided to the Sino-German
                                               CRC 110 ``Symmetries and the Emergence of Structure in QCD'' (Grant No. 11621131001), and the CAS-TWAS
                                               President's Fellowship for International Ph.D. Students.}\\
        CAS Key Laboratory of Theoretical Physics, Institute of Theoretical Physics,\\
        Chinese Academy of Sciences, Beijing 100190, China\\
        University of Chinese Academy of Sciences, Beijing 100049, China\\
        Helmholtz-Institut f\"ur Strahlen- und Kernphysik, Universit\"at Bonn, D-53115 Bonn, Germany\\
        E-mail: \email{naeem@itp.ac.cn}}

\author{Yu Lu\\
        Helmholtz-Institut f\"ur Strahlen- und Kernphysik and Bethe
        Center for Theoretical Physics, \\Universit\"at Bonn,  D-53115 Bonn, Germany\\
        E-mail: \email{luyu@hiskp.uni-bonn.de}}

\author{Bing-Song Zou\\
        CAS Key Laboratory of Theoretical Physics, Institute of Theoretical Physics,\\
        Chinese Academy of Sciences, Beijing 100190, China\\
        University of Chinese Academy of Sciences, Beijing 100049, China\\
        E-mail: \email{zoubs@itp.ac.cn}}

\abstract{ In heavy quarkonia, hadronic transitions serve as an enlightened probe for the structure and help to establish the understanding of light quark coupling with a heavy degree of freedom. Moreover, in recent years, hadronic transitions revealed remarkable discoveries to identify the new conventional heavy quarkonia and extracting useful information about the so called ``XYZ" exotic states.

In this contribution, we present our predictions for heavy quark spin symmetry (HQSS) breaking hadronic transitions of higher $S$ and $D$ wave vector charmonia based on our recently proposed model (inspired by Nambu-Jona-Lasinio (NJL) model) to create light meson(s) in heavy quarkonium transitions. We also suggest spectroscopic quantum numbers $(^{2S+1}L_J)$ for several observed charmoniumlike states. Our analysis indicates that the $Y(4360)$ is most likely to be a $3D$ dominant state.}

\FullConference{XVII International Conference on Hadron Spectroscopy and Structure - Hadron2017\\
		25-29 September, 2017\\
		University of Salamanca, Salamanca, Spain}

\begin{document}

\section{HQSS Violation in Heavy Quarkonium Transitions}

The strong interactions of heavy quarks with light degree of freedom (mesons and gluons) can be described by an effective theory when the scale $1/m_{Q}$ is much smaller than the typical hadronic scale. The leading effective theory is called heavy quark effective theory (HQET), having the feature of heavy quark symmetry (HQS), which is considered invariant under flavor change and spin rotation of the heavy quark~\cite{HQSS}.

Hadronic transitions such as $\psi(2S)\to J/\psi \eta$ or $\psi(4160)\to h_c \eta$ must involve the flip of the initial spin of heavy quark. Such transitions are significantly suppressed as compare to spin keeping transitions such as $\psi(2S)\to J/\psi \pi \pi$. One can estimate the relative sizes of these transitions from the available experimental data, Table~\ref{trans} summarizes the Particle Data Group (PDG) information on these transitions. A better understanding of HQSS violation provides an elegant insight to the dynamics of heavy quark. Since the amplitude of such transitions is $\mathcal M_{\cancel{HQSS}} \propto 1/m_{Q}$, in heavy quark limit ($m_Q \to 0$) this amplitudes vanishes and the spin of heavy quark is conserved. In the actual world
quarks are not infinitely heavy, for the finite mass of the heavy quark ($m_c=1.5$ GeV), one can expect a small breaking of HQSS\footnote{The breaking of HQSS is expected to be enhanced in quarkoniumlike states near the thresholds due to the mixing with heavy meson-antimeson pairs~\cite{Voloshin}.}.

\begin{table}[h]\footnotesize
  \renewcommand\arraystretch{1.5}
  \centering
   \begin{tabular}{ccc}
  \hline\hline
  $\mathcal B[\psi (3686) \to J/\psi \eta]$ & $\mathcal B[\psi (4160) \to h_{c}(1P) \eta]$ & $\mathcal B[\psi (3686) \to J/\psi \pi \pi]$ \\
  \hline
  $(3.36 \pm 0.05)\%$ & $ < 2 \times 10^{-3}$ & $(34.49 \pm 0.30)\%$~~~$\pi^+ \pi^-$\\
  $\Gamma_{\psi(3686)} =  296 \pm 8$ keV& $\Gamma_{\psi (4160)} =  70 \pm 10$ MeV & $(18.17 \pm 0.31)\%$~~~$\pi^0 \pi^0$ \\

  \hline\hline
\end{tabular}
\caption{PDG data on selective HQSS violating and conserving hadronic transitions of $\psi(2S)$ and $\psi(4160)$.}
\label{trans}
\end{table}

\section{Theoretical Developments for Hadronic Transitions}

Soon after the observation of the first charmonium $J/\psi$ and its radial excitation $\psi'\equiv \psi(3686)$, several theoretical formalisms were developed to incorporate the transitions of excited to the lowest charmonium with the emission of light meson(s). The underlying mechanism of such hidden-flavor transitions must govern by strong interactions, and these decays are expected to be OZI allowed. Hence, one can expect these transitions as dominant decay channels for the excited quarkonia below the threshold of disassociation into two open-flavor mesons.

Here we intend to review some pioneer works to describe the hadronic transitions on the footings of the leading theory of strong interactions known as Quantum Chromdynamcs (QCD).

\subsection{QCD Multipole Expansion}

In QCD, the well-established formalism for hadronic transitions is multipole expansion (ME) \cite{QCDME}, which assumes that the hadronic transitions take place due to the intermediate process of gluon emission. These gluons are supposed to be soft, having wavelengths much larger than the size of a heavy quarkonium. These soft gluons further hadronize to light hadron(s) to complete such kinds of hadronic transitions.
Due to the assumption that the intermediate gluons are soft, the emitted light hadron(s) are predominately of the lower momenta. It is very hard to incorporates the transitions of much higher quarkonia in the formalism of QCDME, where we have large decay momenta. For detail discussions and applications of QCDME, we refer the following comprehensive review~\cite{Kuang:2006me}.

\subsection{Effective Theories}

To put hadronic transitions on an effective filed theory (EFT) ground, efforts have been made. The development of heavy meson chiral Lagrangians (HMCL)~\cite{HMCL} is the foremost simplification to QCDME. HMCL serve as an EFT to QCDME in a soft exchange where the gluonic exchanges are predominantly of limited momenta. With the assumptions that (i) the heavy $Q\bar{Q}$ involved in the process is well separated to consider it in a stringlike picture and (ii) the momentum of the emitted light meson is not too large, the HMCL are successful at reproducing the hadronic transitions among lower charmonia.

The transition between two $S$ waves, $S$ to $P$ or $D$ to $S$ wave charmonia with the emission of $\eta$ ($\pi$) might occur through intermediate open-charm contributions.
The formalism which incorporate intermediate heavy mesons within hadrons is referred to as coupled-channel effects (CCEs). CCEs have been taken into account in the QCDME framework~\cite{Zhou:1990ik}.
To investigate the intermediate charmed meson loop effects on $\psi' \to J/\psi \eta (\pi^0)$ decay, a nonrelativistic effective field theory (NREFT) formalism was constructed~\cite{NREFT}. It is noted that if we go to much higher waves e.g., $\psi(nS)$ or $\psi((n-1)D)$ with $n=4,5,6,\ldots$, the decay momentum is not so small, as it lies in the relativistic regime; hence, the NREFT formalisms are not very suitable for studying hadronic transitions of higher charmonia.

The experimental status of the spectrum of higher vector charmonium(like) states is very rich now and several precise measurements have been recorded for their hadronic transitions~\cite{Patrignani:2016xqp}. To describe the observed transitions of higher $c\bar{c}$ systems there is a potential need for a theoretical model which can predict the transitions in the high momentum regime and help to identify the missing higher states through their hidden-flavor decays. We try to fulfill this need by modeling hadronic transitions of higher vector charmonia. Our proposed model is away from all the assumptions of HMCL [(i) and (ii)] and QCDME, and useful to predict the transitions involving much large momenta.

\section{Our Effective Model and Results}

We model the coupling of the light scalar and pseudoscalar meson with the charm quark~\cite{Anwar:2016mxo}. The effective Lagrangian of our model contains both the scalar and pseudoscalar interactions as present in the NJL model. The effective Lagrangian of our proposed model can be written as
\be
\mathcal{L}_I=g(\bar{\psi} \psi <\sigma>+\bar{\psi} i\gamma^5 \psi <\eta>),
\ee
\begin{figure}[h]
  \centering
  \includegraphics[width=0.6\textwidth]{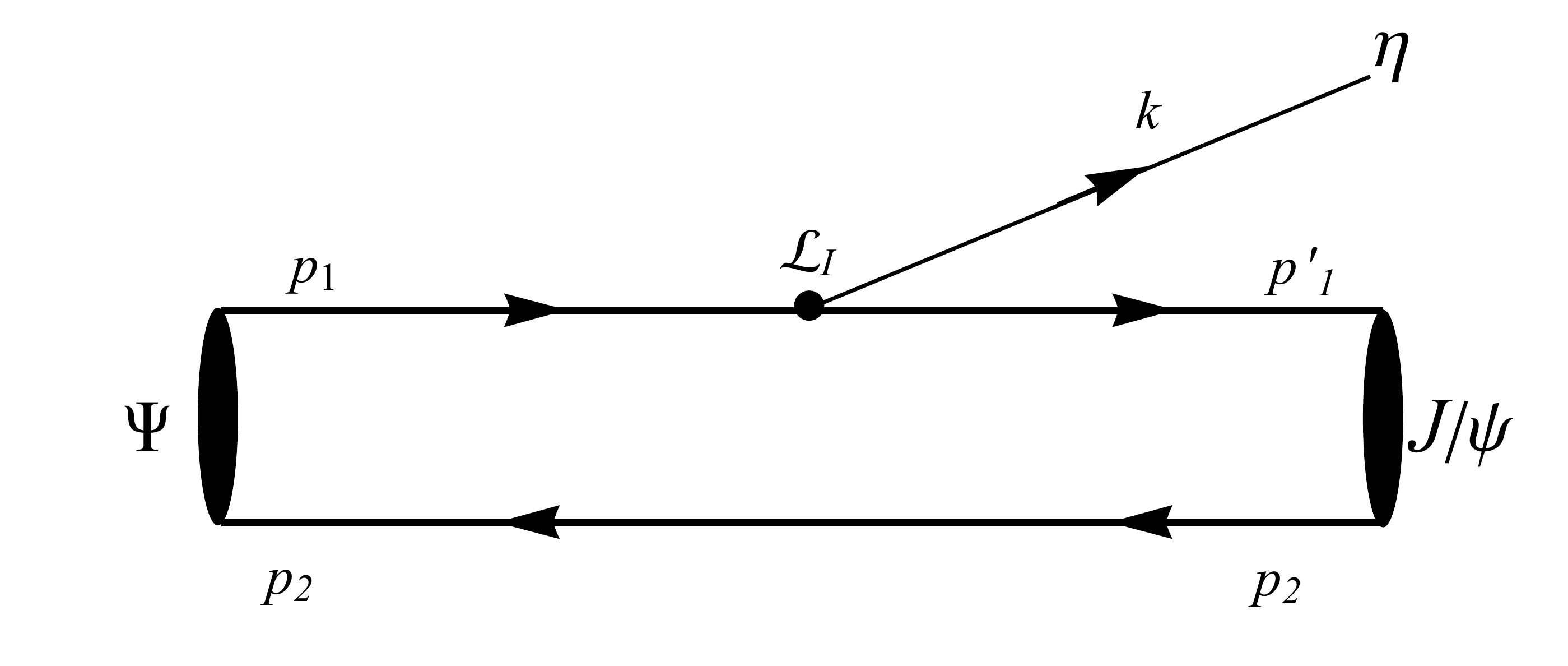}\\
  \caption{Quark level description of higher vector charmonia decaying into $J/\psi \eta$.}
 \label{fyn1}
\end{figure}
where $g$ is the overall coupling strength, $\psi$ is the heavy quark field, and $<\sigma>$ and $<\eta>$ are $SU(3)$ singlet scalar and pseudoscalar meson, respectively. The Lagrangian $\mathcal{L}_I$ allows the coupling of the (anti)quark line only to a scalar or isospin singlet pseudoscalar. The possible Feynman diagram for the process $\Psi \to J/\psi \eta$ is shown in Fig.~\ref{fyn1}. The decay width and transition amplitude of $\Psi\to J/\psi \eta$ (where $\Psi$ is $S$ or $D$ wave higher vector charmonia) is given as
\be
\Gamma_{A\to BC}=2 \pi k \frac{E_B E_C}{m_A} \sum_{m_{J_B},m_{J_C}} \int d\Omega_B\vert \helicityAmp \vert^2,
\label{width}
\ee
\be
\helicityAmp =g \frac{i}{2m_c} \int d^3 p_1
\phi_A(\vec{p}_1) \phi_B^*(\vec{p}_1- \frac{1}{2} \vec{P}_B) \langle 1' \vert \vec{\sigma} \vert 1 \rangle \cdot (\vec{p}_1-\vec{p'}_{1}) \cdot \langle 2 \vert \delta_{ss'} \vert 2' \rangle .
\label{fianl:amp}
\ee
with $E_{B}=\sqrt{m_B^2+k^2}$, $E_{C}=\sqrt{m_C^2+k^2}$ and $k=\sqrt{[m_{A}^2-(m_B-m_C)^2][m_{A}^2-(m_B+m_C)^2]}/2m_A$; $m_c$ is the mass of charm quark. To evaluate the transition amplitude we use simple harmonic oscillator (SHO) wavefunctions. The model parameters are given in Table~\ref{paraTab}.

We consider $J^{PC} = 1^{--}$ charmonia as admixture of $S$ and $D$ waves by adopting the well-established formalism of $S-D$ mixing based on reproducing the dielectric decay widths to deduce the mixing angle~\cite{SDmixing}. For an idea of the parameter dependence of our model, we suggest interested readers to see our parametric plots~\cite{Anwar:2016mxo}.
\begin{table}[h]
  \renewcommand\arraystretch{1.5}
  \centering
\begin{tabular}{cccc}
  \hline\hline
  $m_c=1.5$~GeV  & $\beta=0.40$~GeV & $g=0.80$ & $|\theta|=13\degree$ \\
  \hline\hline
\end{tabular}
\caption{The parameters used in our calculation.}
\label{paraTab}
\end{table}

\subsection{Results and Predictions for $\Gamma(\Psi\to J/\psi \eta)$ and $\Gamma(\Psi \to h_{c}\eta$)}

\begin{figure}[h]
  \centering
  \includegraphics[width=0.8\textwidth]{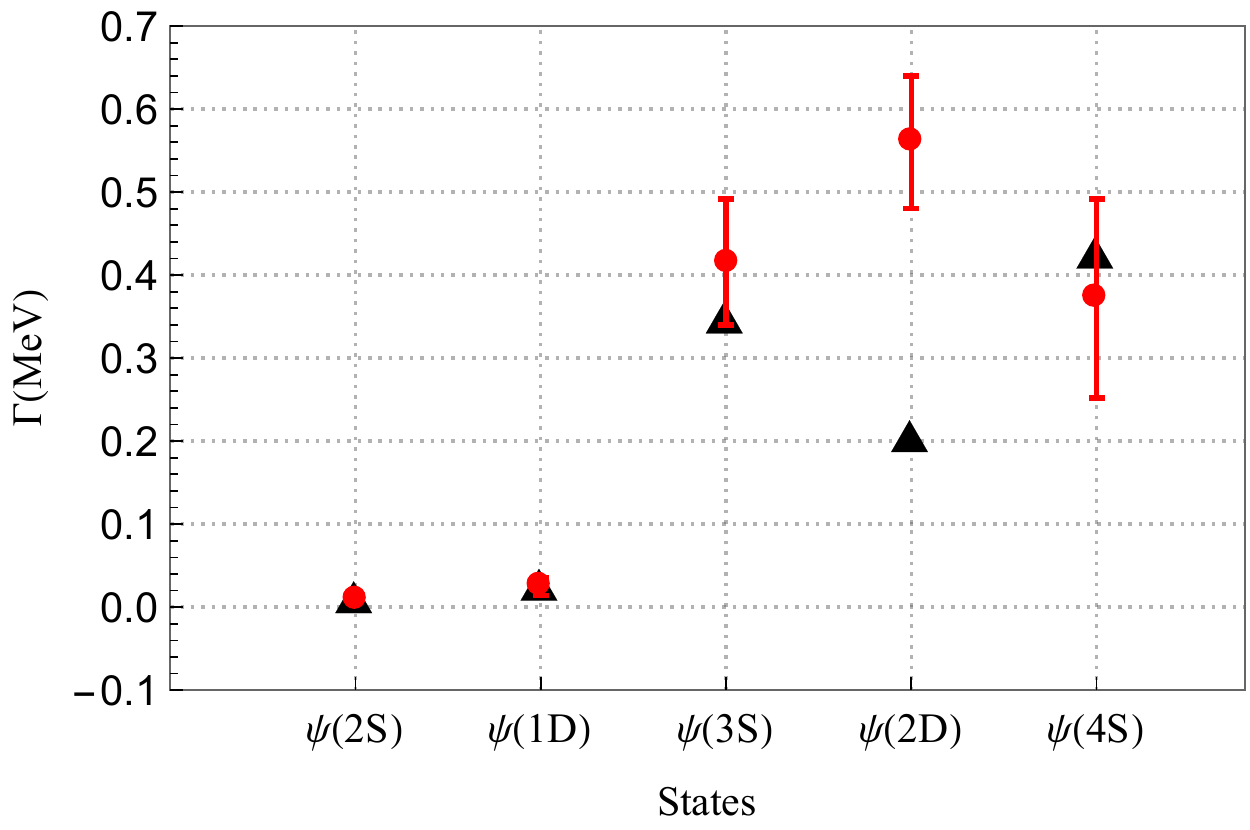}\\
  \caption{Decay width of different higher vector charmonium to $J/\psi \eta$. Red filled circles with error bars are the experimental values taken from PDG and the black triangles are our theoretical predictions. For the case of $\psi(2D)$ and $\psi(4S)$ we only have the upper limits.}
 \label{WidthPlot}
\end{figure}

A graphical presentation of our results for $\Gamma(\Psi\to J/\psi \eta)$ is shown in Fig.~\ref{WidthPlot}. We fit the overall coupling $g$ from the process $\psi' \to J/\psi \eta$, and then allow it to reproduce the similar decays of next excited charmonia. We get quite impressive agreement with the experimental data. Although there exist only upper limits for the $\psi(4160)\to J/\psi \eta$ and $\psi(4415)\to J/\psi \eta$ decay processes, our computed decay width for the former decay process lies within this limit, while for the latter process our predicted width is slighter larger than the central value. It is worthy noting that the experimental value of $\Gamma({\psi(4415) \to J/\psi \eta}$) has large statistical errors. Considering this error range, our prediction in this case still lies within the upper limit.

The threshold for the decay process $\Psi \to h_{c}(1P) \eta$ is $4073$ MeV. The $\psi(4160)$ is the first state which can decay into this final state. The HQSS violating transition $\psi \to h_{c}(1P) \eta$ must requires the spin flip and expected to be significantly suppressed. The observed ratio $\Gamma(\Psi \to J/\psi \eta)/\Gamma(\Psi \to J/\psi \pi \pi)$ is fully consistent with the theoretical predictions~\cite{Voloshin}. It has been argued in~\cite{Guo:2010zk} that the CCEs due to intermediate charmed mesons are quite small for this transition. Our predicted width of $\psi(4415)\to h_{c}(1P) \eta$ is of the same order of magnitude as $\psi(4160)\to h_{c}(1P) \eta$:
\be
\frac{\Gamma(\psi(4160)\to h_{c}(1P) \eta)}{\Gamma(\psi(4160)\to J/\psi \eta)}=7.887\times 10^{-2},~~~
\frac{\Gamma(\psi(4415)\to h_{c}(1P) \eta)}{\Gamma(\psi(4415)\to J/\psi \eta)}=6.736\times 10^{-2}.
\label{hc2D}
\ee

We also give the initial mass dependent decay width of $\psi(nS/(n-1)D) \to J/\psi \eta$ and $h_{c}(1P)\eta$ with $(n=3,4,5,6)$, both for the pure $S$ and $D$ wave and for the standard $S-D$ mixing case~\cite{Anwar:2016mxo}.

\subsection{Study of $\textbf{\textit{Y}}\to J/\psi \eta$ and $\textbf{\textit{Y}}\to h_c \eta$}

Despite the fact that charmoniumlike vector states do not decay into open-charm channels, it would be interesting to study their hidden-charm strong decays. By assuming $Y(4360)$ and $Y(4660)$ as $\psi(3^3D_1)$ and $\psi(5^3S_1)$ dominant states, respectively, we give our predictions for $Y(4360)\to J/\psi \eta$ and $Y(4660)\to J/\psi \eta$. Because only experimental upper limits~\cite{Patrignani:2016xqp} exists for the product of the branching fraction $\mathcal B(Y\to J/\psi \eta)$ and $\Gamma_{e^+ e^-}(Y)$ for $Y(4360)$ and $Y(4660)$, we use average values of the $\Gamma_{e^+ e^-}(Y)$ of available theoretical predictions.

\begin{table}[h]\footnotesize
\renewcommand\arraystretch{1.5}
  \centering
\begin{tabular}{cccc|ccc|ccc}
  \hline\hline
   & & & &\multicolumn{3}{c|}{$\Gamma^{\textrm{th}}_{Y \to J/\psi \eta}$} & \multicolumn{2}{c}{$\Gamma^{\textrm{exp}}_{Y \to J/\psi \eta}$} & \\\cline{5-10}
  State& $n ^{2S+1}L_{J}$ &$\Gamma_{\textrm{total}}$ &$\mathcal B(Y\to J/\psi \eta)$ &$\theta=0\degree$ & $\theta=13\degree$& $\theta=34\degree$ & $\theta=0\degree$  & $\theta=34\degree$ \\
  \hline
  $Y(4360)$ & $3 ^3D_1$  & $ 74\pm 18$~\cite{Patrignani:2016xqp}  & $\frac{6.8}{\Gamma_{e^+ e^-}}$~\cite{Patrignani:2016xqp} &  0.047 & 0.016& $1.0\times 10^{-3}$ & $<0.963$ &$<0.799$
  \\
  $Y(4390)$& $3 ^3D_1 $  & $139.5\pm16.1$~\cite{BESIII:2016adj} & $-$ &$0.083$ & $0.028$ &$ 1.6 \times 10^{-3}$  & $-$  & $-$ &  \\
  $Y(4660)$& $5 ^3S_1 $  & $48\pm 15$~~\cite{Patrignani:2016xqp}  & $\frac{0.94}{\Gamma_{e^+ e^-}}$~\cite{Patrignani:2016xqp} & 0.057 & 0.070  &0.077&  $<0.046$ & $<0.116$\\
  \hline \hline
\end{tabular}
\caption{Predictions for $\Gamma(Y \to J/\psi \eta)$ for the $Y(4360)$, $Y(4390)$, and $Y(4660)$ states. $``-"$ indicates that the experimental data are not available. All the widths are in units of MeV.}
\label{Yassign}
\end{table}

For $Y(4360)$, we give our results in Table~\ref{Yassign} for pure $3D$, small, and large $S-D$ mixing. Our predictions are in agreement with the experimental measurements. We conclude that $Y(4360)$ could be considered as a potential candidate for dominant $3 ^3D_1$ charmonium state. For $Y(4660)$, our prediction for the pure $5S$ case is little above the experimental upper limit which indicates that it might has sizeable $4D$ component. We also present our prediction for $Y(4390)$~\cite{BESIII:2016adj} by assigning it $\psi(3^3D_1)$. To identify this state, measurements on its hadronic branching fraction are required.

The ratio $\frac{\Gamma (Y (4360) \to h_c \eta)}{\Gamma(Y (4360) \to J/\psi \eta)}$ provides another test on the structure of $Y(4360)$. To be a $3 ^3D_1$ state, the order of this ratio should be the same as $\psi(4160)$~\cite{Patrignani:2016xqp}. We will address this in details elsewhere.

\section{Conclusions}

Our study gives an idea of the branching fractions of missing higher vector charmonia into $J/\psi \eta$ and $h_c (1P)\eta$ final states. We suggest that the ongoing (Belle and BESIII) and forthcoming ($\bar{\textrm{P}}$ANDA and BelleII) experiments should look for suggested unobserved decay channels to find $J^{PC}=1^{--}$ higher charmonia. Our estimate of $\eta$ transition branching fractions for $Y(4360)$ by assuming it as $3 ^3D_1$ is consistent with experimental data. Hence, we argued that the $Y(4360)$ can be considered as a potential candidate for the $3 ^3D_1$ charmonium. Assuming $Y(4660)$ to be $5 ^3S_1$, the predictions are consistent within the experimental upper limit. We hope that our predictions provide useful references to search and better understanding of higher charmonia.

\end{document}